\documentclass[preprint,12pt,nofootinbib]{revtex4-1}
\usepackage[paper=letterpaper,margin=1.5in]{geometry}

\usepackage{graphicx}\graphicspath{%
{graph/}}%
\usepackage{amssymb,amsmath,amsfonts}
\usepackage{time}
\usepackage{epstopdf}
\usepackage{hyperref}
\hypersetup{
  colorlinks,
  citecolor=magenta,
  linkcolor=blue}
\usepackage[T2A,T1]{fontenc} 
\usepackage[utf8]{inputenc} 
\usepackage{txfonts} 

\newcommand{\p}{{\partial}}

\begin{document}

\begin{flushright}
{\normalsize
SLAC-PUB-16457\\
LCLS-II TN-16-01\\
arXiv submit/1465446\\
January 2016}
\end{flushright}

\vspace{.4cm}

\title{Dechirper Wakefields for Short Bunches \footnote[1]{Work supported by the U.S. Department of Energy, Office of Science, Office of Basic Energy Sciences, under Contract No. DE-AC02-76SF00515
} }

\author{Karl Bane\footnote[2]{kbane@slac.stanford.edu} and Gennady Stupakov}
\affiliation{SLAC National Accelerator Laboratory,Menlo Park, CA 94025}

\begin{center}
\end{center}

\begin{abstract}
In previous work~\cite{surface_imped_flat_geometry} general expressions, valid for arbitrary bunch lengths, were derived for the wakefields of corrugated structures with flat geometry, such as is used in the RadiaBeam/LCLS dechirper. However, the bunch at the end of linac-based X-ray FELs---like the LCLS---is extremely short, and for short bunches the wakes can be considerably simplified. In this work, we first derive analytical approximations to the short-range wakes. These are generalized wakes, in the sense that their validity is not limited to a small neighborhood of the symmetry axis, but rather extends to arbitrary transverse offsets of driving and test particles. The validity of these short-bunch wakes holds not only for the corrugated structure, but rather for any flat structure whose beam-cavity interaction can be described by a surface impedance. We use these wakes to obtain, for a short bunch passing through a
dechirper: estimates of the energy loss as function of gap, the transverse kick as function of beam offset, the slice energy spread increase, and the emittance growth. 
In the Appendix, a more accurate derivation---than is found in~\cite{surface_imped_flat_geometry}---of the arbitrary bunch length wakes is performed; we find full agreement with the earlier results, provided the bunches are short compared to the dechirper gap, which is normally the regime of interest.

\vspace{2mm}
\end{abstract}

\maketitle

\section*{Introduction}\label{sec:0}

The idea of using a corrugated structure as a ``dechirper" in a linac-based X-ray FEL was first proposed in Ref.~\cite{Bane12}. The idea was to use a passive device to remove residual energy chirp in the beam before it enters the undulator for lasing. The original report considered a round structure. However, to allow for adjustability, it was next proposed to use a corrugated structure in flat geometry. But because of an unavoidable quadrupole wake excited in flat geometry, the structure is best built in two halves, with one half rotated by 90 degrees with respect to the other, in order to, in principle, have the quad wake forces cancel.
Such structures have been built by RadiaBeam and tested, first at Pohang~\cite{Emma14} and more recently at the Linac Coherent Light Souce (LCLS)~\cite{LCLSdechirper},\cite{dechirperpaper}. The RadiaBeam/LCLS dechirper is the first one that has been tested at high energies (multi GeV) and short bunch lengths (10's of microns). 

For a nominal set of parameters for the LCLS,
the dominant wavelength of the wakefields $\sim2.5$~mm~(see {\it e.g.}~\cite{Zhang15}). The full bunch length, however, is normally small,  $\lesssim 100$~$\mu$m. Thus, we can expect that a model of the longitudinal (steady-state) wake that assumes that it is constant and equal to its value at the origin can be used to approximate the wake effects of the beam in such machines. The real wake will drop from the origin~(see \cite{Novo15}), and such a model will give a maximum bound of the wake, one that becomes more accurate as the bunch becomes shorter (assuming it is not so short that the transient component of the wake becomes significant). For the transverse wake, equal to zero at the origin, we use an approximation of a linear function with the constant slope equal to that at the origin. 
We will see that, compared to numerically calculating the wakes, we gain by obtaining simple analytical functions that show us the structure of the wakes.

The calculation of wakes of corrugated structures has been performed for round structures, assuming small corrugations and using perturbation methods~\cite{BaneNovo99}, \cite{BaneStupakov00}. The same has been done for flat geometry~\cite{BaneStupakov03}.Time domain simulations have been performed for more accurate results, particularly when the corrugation parameters are not small compared to the dechirper gap~\cite{Cho13}, \cite{Zagorodnov15}, \cite{Novo15}. More recently, a detailed analysis of the RadiaBeam/LCLS dechirper has been performed using field matching methods~\cite{Zhang15}.  
Finally, in Ref.~\cite{surface_imped_flat_geometry}, for the case of flat geometry and assuming the impedance can be characterized by a surface impedance, equations for the {\it generalized} wakefields, valid for arbitrary bunch length, are derived.  By ``generalized" we mean (point charge) wake functions for which the transverse positions of driving and test particles are arbitrary, and are not limited to being near the symmetry plane. 

We begin the present report with the results of~\cite{surface_imped_flat_geometry} and obtain the values at the origin of the (generalized) longitudinal wake and the slope of the transverse wakes. 
That is enough to obtain an approximation to the energy loss and kick experienced by a pencil beam (one with no transverse extent). There is some interest in using a dechirper as a fast kicker, by passing a beam close to one jaw (see {\it e.g.}~\cite{Novo15}, \cite{Alberto}), and the case of the beam near the wall is automatically included.
Then, assuming that the transverse beam sizes are small compared to the gap, we derive perturbation solutions of the wakes, valid over a small neighborhood around the beam centroid. This allows us to find wake effects like slice energy spread increase and projected emittance growth. Next, we briefly investigate the accuracy of our model. We finally end with conclusions. In the Appendix, we give a more accurate derivation, than found in \cite{surface_imped_flat_geometry}, of the flat geometry wakes for arbitrary bunch lengths.

Representative beam and machine parameters for the LCLS, which will be used in example calculations in this report, are given in Table~I. Note that the entire RadiaBeam/LCLS dechirper assembly comprises two 2-m-long sections, one oriented horizontally, the other vertically. The wakes are given here in cgs units; to convert to MKS one merely needs to multiply by $(Z_0c/4\pi)$, with $Z_0=377$~$\Omega$ and $c$ is the speed of light. Wakefield induced energy loss, however, has been converted to MKS units, for convenience.

\begin{table}[hbt]
   \centering
   \caption{Selected beam and machine properties (at the dechirper) used in example calculations. This is a typical combination of parameters found in Ref.~\cite{Zhang15}. The longitudinal bunch distribution is taken to be uniform. In the current lattice, the first (second) dechirper is vertically (horizontally) orientated, with positive (negative) $\alpha$.}
   \begin{tabular}{||l|c|c||}\hline 
        {Parameter name} & {Value}  &  Unit\\ \hline\hline 
      Beam energy, $E$       &6.6  &GeV \\
       Charge per bunch, $Q$       &150  &pC  \\
      Beam current, $I$       &1.5  &kA  \\
       Full bunch length, $\ell$       &30  &$\mu$m \\
      Normalized emittance, $\epsilon_{xn}$ / $\epsilon_{yn}$      &0.77 / 0.39  &$\mu$m  \\
      Beta function, $\beta_{x}$ / $\beta_{y}$      &4.5 / 23.7  &m  \\
      Alpha function, $\alpha_{x}$ / $\alpha_{y}$      &$\pm0.024$ / $\pm1.572$  &  \\
      Beam size, $\sigma_{x}$ / $\sigma_{y}$      &16 / 27  &$\mu$m  \\
      Dechirper half aperture, $a$       &0.7  &mm \\      
      Dechirper section length used, $L$       &2  &m \\
         \hline \hline 
   \end{tabular}
   \label{table1_tab}
\end{table}

\section*{Generalized wakefields for short bunches}

\subsection*{Longitudinal Wake}

\begin{figure}[htb]
\centering
\includegraphics[width=0.6\textwidth, trim=0mm 0mm 0mm 0mm, clip]{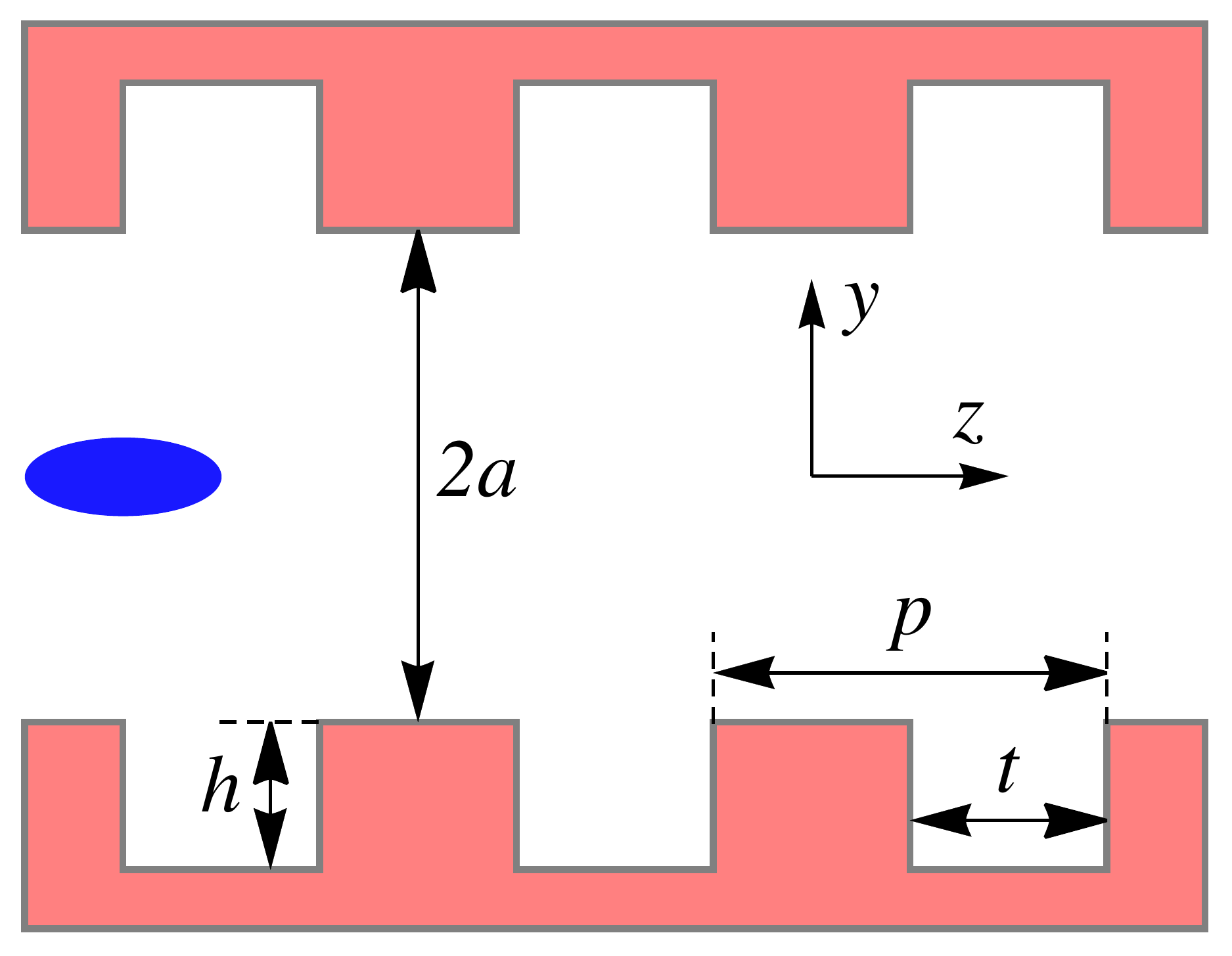}
\caption{Geometry of a horizontal dechirper. A rectangular coordinate system is centered on the symmetry axis of the chamber. The blue ellipse represents an electron beam propagating along the $z$ axis.}
\label{fig:1}
\end{figure}

We consider a horizontally oriented dechirper, of full gap $2a$, with the symmetry plane located at $y=0$, see Fig.~\ref{fig:1}.
In the case of the RadiaBeam/LCLS dechirper, the corrugation parameters are period $p=0.5$~mm, depth $h=0.5$~mm, and longitudinal gap $t=0.25$~mm. The plate width in $x$, $w=12.7$~mm. With the typical gap $a\lesssim$ a few millimeters, the width $w$ does not affect the short-range wakes, and we can let $w\rightarrow\infty$. This is the model we consider and what we designate by the term ``flat geometry."

To calculate both the longitudinal and transverse (point charge) wakes at the origin, $s=0$, ($s$ the distance the test particle trails the driving particle) we note that they are defined by the asymptotic behavior of the impedance at wave number $k\to\infty$. The relation between the longitudinal wake and the impedance is
    \begin{align}\label{eq:3}
    Z_l(k)
    =
    \frac{1}{c}
    \int_{-\infty}^\infty
    ds\,w(s)e^{iks}
    \ ,
    \end{align}
with $w(s)$ the (point charge) wake. Near the origin the wake can be approximated by the step function
\begin{equation}
w(s)= w_0 h(s)\ , 
\end{equation}
with $w_0$ the value of the wake at $s\to +0$; $h(s)=0$ for $s<0$, 1 for $s>0$ . Substituting the approximated wake into~\eqref{eq:3} and performing the integration we find that the asymptotic form of the impedance $Z_{la}(k)$
    \begin{align}\label{eq:4}
    Z_{la}(k)
    =
    \frac{w_0}{c}
    \int_{0}^\infty
    ds\,h(s)e^{iks}
    =
    \frac{iw_0}{kc}
   \  ,
    \end{align}
where we have neglected the contribution to the integral at the upper limit. At high frequencies the leading order term of the impedance is imaginary, with $Z_{la}\sim i/k$. Thus, 
\begin{equation}
w_0=-ikcZ_{la}(k)=-ic\lim_{k\to\infty} kZ_{l}(k)\label{w0_eq}
\end{equation}
which is a positive constant.

In general, as we shall see below, the wake at the origin $w_0$ is a function of both driving and test particle transverse positions, $(x_0,y_0)$ and $(x,y)$. For the {\it bunch wake} $W_\lambda(x,y,s)$---{\it i.e.} the wake seen by a test particle within a bunch---it is often sufficent to consider a pencil beam, that is one without transverse extent. For this case, for a short bunch, the wake is obtained by setting $x_0=x$, $y_0=y$, and performing the convolution
\begin{equation}
W_\lambda(y,s)=-\int_0^\infty w(s')\lambda(s-s')\,ds'\approx -{w_0(x_0=x,y_0=y)}\int_0^s\lambda(s')\,ds'\ ,\label{wlambda_eq1}
\end{equation}
with $\lambda(s)$ the longitudinal bunch distribution. (Note, a pencil beam wake has no $x$ dependence.) When a transverse extent needs to be included, for example when calculating the slice energy spread increase or the emittance growth, we need to average over $x_0$ and $y_0$ to find the bunch wake:
\begin{eqnarray}\label{eq:6}
W_\lambda(x,y,s)&=&-\int\lambda_x(x_0)\,dx_0\int\lambda_y(y_0)\,dy_0\int_0^\infty w(x_0,y_0,x,y,s')\lambda(s-s')\,ds'\nonumber\\
&&\approx -\int dx_0\lambda_x(x_0)\int dy_0\lambda_y(y_0)\,{w_0(x_0,y_0,x,y)}\int_0^s\lambda(s')\,ds'\ ,
\end{eqnarray}
with $\lambda_x(x)$, $\lambda_y(y)$, the transverse bunch distributions. In Eq.~\ref{eq:6} we have assumed that the 3D distribution function of the bunch can be represented as a product $\lambda_x(x)\lambda_y(y)\lambda(s)$.

In the LCLS the longitudinal bunch shape is rather uniform; for such a case
the pencil beam wake
\begin{equation}
W_\lambda(s)\approx- w_0\frac{s}{\ell}\ , 
\end{equation}
with $\ell$ the full bunch length. The loss factor $\varkappa$, minus the average of the bunch wake, is simply given by $\varkappa=\frac{1}{2} w_0$.

To obtain the high frequency longitudinal impedance for the dechirper, we start from the general impedance expression~\cite{surface_imped_flat_geometry} 
\begin{equation}
 Z_l(k)=\frac{2\zeta}{c}\int_{-\infty}^\infty dq\,q\,\mathrm{csch}^3(2qa) f(q,y,y_0) e^{-iq(x-x_0)}\ ,\label{tildeZl2_eq}
\end{equation}
where $\zeta$ is the surface impedance, $a$ is the half-gap of the dechiper, and the driving and test particles have transverse positions of $(x_0,y_0)$ and $(x,y)$, respectively. The function $f=N/D$, with
\begin{eqnarray}\label{eq:9-1}
N=q(\cosh[q(2a-y-y_0)]&&-2\cosh[q(y-y_0)]+\cosh[q(2a+y+y_0)])\nonumber\\
&&\!\!\!\!\!\!\!\!\!\!\!\!\!\!\!\!\!\!-ik\zeta(\sinh[q(2a-y-y_0)]+\sinh[q(2a+y+y_0)])\ ,\nonumber\\
D=[q\,\mathrm{sech}(qa)-ik\zeta&&\mathrm{csch}(qa)][q\,\mathrm{csch}(qa)-ik\zeta\mathrm{sech}(qa)]\ .
\end{eqnarray}

Eqs.~\ref{tildeZl2_eq} and~\ref{eq:9-1} are derived in Ref.~\cite{surface_imped_flat_geometry} using an approach that does not reveal all the constraints required for its validity. In the Appendix we perform a more accurate derivation of the general wake solution for the dechirper than is found in Ref.~\cite{surface_imped_flat_geometry}. We find that Eqs.~\ref{tildeZl2_eq} and~\ref{eq:9-1} are valid provided that the bunch length over half gap, $\sigma_z/a$, is small, which is indeed the parameter regime of interest.

We now take the limit of large $k$ in Eq.~\ref{tildeZl2_eq} by neglecting terms on the order of $1/k$ and find
    \begin{equation}\label{eq:1}
    Z_{la}(k)
    =
    \frac{2i}{ck}\int_{-\infty}^\infty dq\,q\,e^{-iq(x-x_0)}\, 
    \frac{\sinh[q(2a-y-y_0)]+\sinh[q(2a+y+y_0)]}
    {\mathrm{sinh}^3(2qa)\mathrm{csch}(qa)\mathrm{sech}(qa)} 
   \  .
    \end{equation}
The integral can be performed analytically, and when combined with Eq.~\ref{w0_eq}, we find that
    \begin{eqnarray}\label{eq:5}
    w_0
    &=&
    \frac{\pi ^2}{4 a^2}{\rm Re}\left[
    {\rm sech} ^2\left(\frac{\pi [(x-x_0)+i (y+y_0)]}{4 a}\right)\right]\nonumber\\
&=& \frac{\pi ^2}{2 a^2}\frac{1+\cosh\left[\frac{\pi(x-x_0)}{2a}\right]\cos\left[\frac{\pi(y+y_0)}{2a}\right]}{\left(\cosh\left[\frac{\pi(x-x_0)}{2a}\right]+\cos\left[\frac{\pi(y+y_0)}{2a}\right]\right)^2}\label{w0_big_eq}
   \ .
    \end{eqnarray}
Note that the second form of Eq.~\ref{w0_big_eq}, for the special case $x_0=y_0=0$, can be found derived in Ref.~\cite{Baturin15}. 
We note that the Laplacian applied to Eq.~\ref{eq:5} equals 0, as it must for wakefields~(see {\it e.g.} Ref.~\cite{Klatt86}). In addition, if we exchange $(x_0,y_0)$ and $(x,y)$, $w_0$ remains unchanged, which is required of a structure with flat geometry~\cite{Zagorodnov15}.
We see, in addition, that, when both particles are on-axis, we obtain a minimum value of $w_0=\pi^2/4a^2$; the same result is obtained if both particles have the same $x$ position with $y=-y_0$. 

Note that the surface impedance $\zeta$ does not enter into our final result, Eq.~\ref{eq:5}. Thus, this result is valid not only for the dechirper, but rather for any structure with flat geometry whose boundary condition can be described by a surface impedance. For example, it is also valid for a resistive wall impedance~\cite{AChao}, for the impedance of a perfectly conducting pipe with shallow corrugations~\cite{Stupakov00}, and for a disk-loaded accelerating structure at high frequencies~\cite{Stupakov95}. 

For a pencil beam ($x_0=x,y_0=y$), Eq.~\ref{eq:5} becomes
\begin{equation}
w_0=\frac{\pi ^2}{ 4a^2}\sec^2\left(\frac{\pi y}{2a}\right)\ .
\end{equation}
The minimum loss is for a beam on axis, when $w_0=\pi^2/4a^2$. There is some interest in using a dechirper as a fast kicker, by passing a beam close to one jaw; if the beam is near the wall, {\it i.e.} $y=a-d$ with $d/a$ small, then $w_0\approx 1/d^2$.
The average energy loss (in MKS units, in [eV]) of a pencil beam is
\begin{equation}
\left( E_w\right)_{av}=\frac{Z_0ceQL}{8\pi}w_0=\frac{\pi}{32a^2}Z_0ceQL\sec^2\left(\frac{\pi y}{2a}\right)\ ,
\end{equation}
with $Q$ the bunch charge and $L$ the dechirper length. The induced chirp is twice this value, divided by the total bunch length $\ell$, with the tail of the beam losing the most energy. 
For a beam on axis, for the representative parameters of Table~I, $(E_w)_{av}=6.8$~MeV, with the bunch tail losing twice this amount.

\subsection*{Transverse Wake}

In the transverse case, the impedance  $Z_y(k)$ is connected to the wake $w_y(z)$ by
    \begin{align}\label{eq:8}
    Z_y(k)
    =
    -
    \frac{i}{c}
    \int_{-\infty}^\infty
    ds\,w_y(s)e^{iks}
    \ ,
    \end{align}
and the same kind of formula applies to the impedance in $x$, $Z_x(k)$.
Near the origin the (point charge) wake can be approximated by 
\begin{equation}
 w_y(s)\approx w'_{y0} s\,h(s)\ , 
\end{equation}
with $w'_{y0}$ the value of the slope of the wake at $s\to +0$. Substituting this approximation into~\eqref{eq:8}, we find the asymptotic impedance
    \begin{align}\label{eq:9}
    Z_{ya}(k)
    =
    -
    \frac{iw'_{y0}}{c}
    \int_{0}^\infty
    ds\,se^{iks}
    =
    \frac{iw'_{y0}}{ck^2}
    \ ,
    \end{align}
where again we have neglected the contribution to the integral at the upper limit. Thus, from the high frequency impedance we obtain
\begin{equation}
w_{y0}'=-ik^2cZ_{ya}(k)=-ic\lim_{k\to\infty} k^2Z_{ya}(k)\ ,\label{w0p_eq}
\end{equation}
which is a positive constant. 

The  
vertical {\it bunch wake} $W_{\lambda y}(s)$ for a short bunch, pencil beam is obtained by the convolution
\begin{equation}
W_{\lambda y}(y,s)=\int_0^\infty w_y(s')\lambda(s-s')\,ds'\approx {w_{y0}'(x_0=x,y_0=y)}\int_0^s\lambda(s')s'\,ds'\ ,
\end{equation}
and there is no horizontal wake force.
(For a bunch with a transverse distribution, averaging needs to be performed over $x_0$, $y_0$, just like in the longitudinal case discussed above.)
For a uniform bunch distribution, the bunch wake 
\begin{equation}
W_{\lambda y}(y,s)\approx w_{y0}'\frac{s^2}{2\ell}\ ,
\end{equation}
 with $\ell$ the full bunch length. The kick factor $\varkappa_y$, the average of the bunch wake, is simply given by $\varkappa_y=\frac{1}{6}w_{y0}'\ell$.

To calculate the slope of the transverse wake at the origin for the dechirper we begin with the Panofsky-Wenzel theorem~\cite{PanofskyWenzel56}, $Z_y=(1/k)(\partial Z_l/\partial y)$. Combining with Eqs.~\ref{w0_eq}, \ref{w0p_eq}, we obtain
\begin{equation}
w_{y0}'=\frac{\partial w_0}{\partial y}\ .\label{wy0p2_eq}
\end{equation} 
The final answer is obtained by substituting from Eq.~\ref{eq:5}.
For the wake in the $x$ direction, in the same manner, one solves equations that correspond to Eqs.~\ref{eq:8}--\ref{wy0p2_eq}. In compact form, one can write the final answer as
\begin{equation}
\bar w_0'\equiv w_{x0}'+i \,w_{y0}'={\rm Re}(f)+i\,{\rm Im}(f)\ ,\label{w0pb_eq}
\end{equation}
with
    \begin{equation}\label{eq:7-1}
    f
     =-
     \frac{\pi ^3}{8a^3}
     \tanh \left(\frac{\pi [(x-x_0)-i
     \left(y+y_0\right)]}{4 a}\right)
     {\rm sech} ^2\left(\frac{\pi [(x-x_0)-i
     \left(y+y_0\right)]}{4 a}\right)
    \  .
     \end{equation}
Again we find that $\zeta$ does not enter in the final result, and that the wakes $w_{x0}'$, $w_{y0}'$, satisfy Laplace's equation. If we exchange $(x_0,y_0)$ and $(x,y)$ then $(w'_{x0},w'_{y0})\rightarrow(-w'_{x0},w'_{y0})$.
For the special case $x=x_0$,
\begin{equation}
 w_{x0}'=0\ ,\quad w_{y0}'
     =
     \frac{\pi ^3}{8a^3}
     \tan \left(\frac{\pi(y+y_0)}{4 a}\right)
     \sec^2\left(\frac{\pi 
     (y+y_0)}{4 a}\right)
    \  .
     \end{equation}
For a pencil beam near the wall, {\it i.e.} for $x_0=x$, $y_0=y$, and $y=a-d$ with $d/a$ small, then $w_{y0}'\approx 1/d^3$.

For a pencil beam, the average vertical kick experienced by the beam,
\begin{equation}
\left(y'_w\right)_{av}=\frac{Z_0c}{24\pi}\left(\frac{eQL\ell}{E}\right)w_{y0}'=\frac{\pi^2}{192a^3}Z_0c\left(\frac{eQL\ell}{E}\right)\sec^2\left(\frac{\pi y}{2a}\right)\tan\left(\frac{\pi y}{2a}\right)\ .
\end{equation}
For the parameters in Table~I, near the axis, the average vertical kick is $(y_w')_{av}/y=52$~$\mu$r/mm; the kick at the bunch tail is three times larger.

%
\section*{Wakes for beams with small transverse size}\label{sec:II}
%

\subsection*{Longitudinal Wake}

For the purpose of computing the longitudinal wake effect on slice energy spread, we need to know the wake in a small neighborhood of transverse space. In this report, for simplicity, let us assume that the transverse bunch distributions (in $x$ and $y$) are symmetric about a centroid. Consider that a beam is centered at coordinates $(0,y_c)$, where a driving and test particle have respectively coordinates 
\begin{equation}
(x_0,y_0)=(x_0,y_c+\Delta y_0)\ ,\quad\quad (x,y)=(x,y_c+\Delta y)\ ,\label{xx0_eq}
\end{equation}
with $\Delta y_0/a$, $\Delta y/a\ll1$. 
Then Eq.~\ref{eq:5} can be expanded as
\begin{eqnarray}
w_0&\approx&\frac{\pi^2}{4a^2}\sec^2\left(\frac{\pi y_c}{2a} \right)\bigg(1+\frac{\pi}{2a}\tan\left(\frac{\pi y_c}{2a} \right)\left(\Delta y_0+\Delta y\right)\nonumber\\
&&\quad+\frac{\pi^2}{16a^2}\sec^2\left(\frac{\pi y_c}{2a} \right)\left[2-\cos\left(\frac{\pi y_c}{a}\right)\right]
\left[(\Delta y_0+\Delta y)^2-(x_0-x)^2\right]\bigg)\ .\label{w0app_eq}
\end{eqnarray}
Here we have kept terms to second order in $x_0/a$, $x/a$, $\Delta y_0/a$, $\Delta y/a$. 
With the particles near the axis ($y_c=0$), we have
\begin{equation}
w_0=\frac{\pi^2}{4a^2}\bigg(1+\frac{\pi^2}{16a^2}
\left[( y_0+ y)^2-(x_0-x)^2\right]\bigg)\ .\label{w0_axis_eq}
\end{equation}
With the particles near the wall ($y_c=a-d$, with $d/a$ small), and with both particles' offsets being small compared to $d$, we have
\begin{equation}
w_0=\frac{1}{d^2}\bigg(1+\frac{( \Delta y_0+ \Delta y)}{d}+\frac{3}{4d^2}
\left[( \Delta y_0+ \Delta y)^2-(x_0-x)^2\right]\bigg)\ .\label{w0_wall_eq}
\end{equation}

\subsection*{Slice Energy Spread}

Since the longitudinal wake depends on the transverse offset of the driving and test particles, for a distribution of particles with transverse extent, the wake will increase the slice energy spread. For simplicity let us assume that the beam is bi-Gaussian in $x$ and $y$, with rms respectively of $\sigma_x$ and $\sigma_y$, and with a uniform distribution in $s$. Then the rms energy spread of slice $s$ is given by $(W_\lambda)_{rms}^2=\langle W_\lambda^2\rangle-\langle W_\lambda\rangle^2$, where averaging means to integrate over the Gaussian distributions in $x$, $\Delta y$. For $W_\lambda$ we consider the expression given in Eq.~\ref{w0app_eq} averaged over $x_0$ and $y_0$ according to Eq.~\ref{eq:6}. We note that two terms will contribute to $(W_\lambda)_{rms}^2$: a constant times $\left(\Delta y_0+\Delta y\right)_{rms}^2/a^2$, and another constant times $[(\Delta y_0+\Delta y)^2-(x_0-x)^2]_{rms}^2/a^4$. Performing the averaging over the transverse distributions, we find that the rms of these combination of parameters are
\begin{eqnarray}
&&\left(\Delta y_0+\Delta y\right)_{rms}^2=\langle(\Delta y_0+\Delta y)^2\rangle=\sigma_y^2\ ,\nonumber\\ 
&&\left[(\Delta y_0+\Delta y)^2-(x_0-x)^2\right]_{rms}^2  =2\left(\sigma_x^4+\sigma_y^4\right)\ .\label{rms_eq}
\end{eqnarray}

Now let us consider the rms slice energy spread induced in the beam in the MKS system (in eV) $(\bar E_w)_{rms}(s)$, assuming that the beam is sufficiently short with a uniform longitudinal distribution. In the general case (Eq.~\ref{w0app_eq}), with $y_c/a$ not small, $y_c\gg \sigma_y$, (and assuming $\sigma_x/a$, $\sigma_y/a$ small), we find that
\begin{equation}
\left(\bar E_w\right)_{rms}(s)=\left(\frac{Z_0c}{4\pi}\right){eQL}(w_0)_{rms}\left(\frac{s}{\ell}\right)=\frac{\pi^2}{32}\sec^2\left(\frac{\pi y_c}{2a} \right)\tan\left(\frac{\pi y_c}{2a} \right)Z_0c\left(\frac{eQLs}{a^3\ell}\right)\sigma_y\ ,
\end{equation}
with $\ell$ the full bunch length. 
With the beam centered on axis, we use Eqs.~\ref{w0_axis_eq}, \ref{rms_eq}, to obtain
\begin{equation}
\left(\bar E_w\right)_{rms}(s)=\frac{\sqrt{2}\pi^3}{256}Z_0c\left(\frac{eQLs}{a^4\ell}\right)\left(\sigma_x^4+\sigma_y^4\right)^{1/2}\ .
\end{equation}
With the beam near the wall ($y_c=a-d$, with $d/a$ small)
\begin{equation}
\left(\bar E_w\right)_{rms}(s)=\frac{Z_0c}{4\pi}\left(\frac{eQLs}{d^3\ell}\right)\sigma_y\ .
\end{equation}

If we consider just one 2-m-long half of dechirper (and assume the other half is completely opened up and not affecting the beam) and take the beam on axis with representative parameters and $a=0.7$~mm (see Table~I), we find the rms energy spread induced at the tail of the bunch is $(\bar E_w)_{rms}(\ell)=18.7$~keV. Note that if we, instead, were to move the beam near only one jaw, at a distance $d=0.7$~mm, and move the other jaw far away, we obtain $(\bar E_w)_{rms}(\ell)=210$~keV.

\subsection*{Transverse Wake}

Again consider the case when a driving and test particle coordinates are given by Eq.~\ref{xx0_eq} with $\Delta y_0/a$, $\Delta y/a\ll1$. Then, keeping leading terms in Eq.~\ref{w0pb_eq} we obtain
\begin{equation}
w_{x0}'=-w_{q0}'(x-x_0)\ ,\quad
w_{y0}'=w_{d0}' +w_{q0}'\Delta y\label{wxy0pexp_eq}\ ,
\end{equation}
with 
\begin{eqnarray}
w_{q0}'&=&\frac{\pi^4}{32a^4}\left[2-\cos\left(\frac{\pi y_c}{a} \right) \right]\sec^4\left(\frac{\pi y_c}{2a} \right)\ ,\nonumber\\
w_{d0}'&=&\frac{\pi^3}{8a^3}\sec^2\left(\frac{\pi y_c}{2a} \right)\tan\left(\frac{\pi y_c}{2a} \right) +w_{q0}'\Delta y_0\ .
\ \label{wq0_eq}
\end{eqnarray}
Note the wake is of the form
\begin{equation}
w_x=-w_{yq}(x-x_0)\ ,\quad\quad w_y=w_{yd}+w_{yq}\Delta y\ ,
\end{equation}
which contains only two independent wake components, the (vertical) dipole ($w_{yd}$) and quad ($w_{yq}$) wake functions; such a form is required for the local behavior of wakes in flat geometry.
The quad wake term in Eq.~\ref{wxy0pexp_eq}, $w_{q0}'$, gives a kick that is focusing in $x$ (kicking toward the driving particle) and  defocusing in $y$ (kicking away from the axis). Its amplitude gives the inverse focal length of the quad kick. 


For particles with small offsets from the axis ($y_c=0$), we have
\begin{equation}
 w_{x0}'=-\frac{\pi^4}{32a^4}(x-x_0)\ ,\quad w_{y0}'=\frac{\pi^4}{32a^4}(y_0+y)\ .
\end{equation}
Close to the wall ($y_c=a-d$, with $d/a$ small, and with particle offsets small compared to $d$) we have
\begin{equation}
 w_{x0}'=-\frac{3}{2d^4}(x-x_0)\ ,\quad w_{y0}'=\frac{1}{d^3}+\frac{3}{2d^4}(\Delta y_0+\Delta y)\ .\label{wq0_wall_eq}
\end{equation}

\subsection*{Emittance Growth}

For a beam not centered on-axis, the dipole wake will give a kick to the beam's slice centroids, one that increases from head to tail of the bunch and results in projected emittance growth. The quad wake---even for a beam centered on axis---will focus (defocus) in $x$ ($y$) by an amount that increases from the head to tail of the bunch, and also results in projected emittance growth. Note, however, that the slice emittance is not affected by the wakes, at least not for the wakes taken to this order.

Let us begin with considering the quad wake. The inverse focal length of the quad kick in a short, uniform bunch is given by
\begin{equation}
f_q^{-1}(s)=k_q^2(s)L=\left(\frac{Z_0c}{8\pi}\right)\left(\frac{eQL}{E\ell}\right)w_{q0}'s^2\ ,\label{fm1_eq}
\end{equation}
with $k_q(s)$ the effective quad strength, and with the quad wake component $w_{q0}'$ given, in general, by Eq.~\ref{wq0_eq}. The quad wake effect can be considered significant whenever $\beta f_q^{-1}(\ell)\gtrsim1$, with $\beta$ representing the larger lattice beta function at the dechirper. For the parameters in Table~I, with the beam centered on the axis ($y_c=0$ in Eq.~\ref{wq0_eq}), we find that $\beta f_q^{-1}(\ell)=0.35$ (1.8) in $x$ ($y$). Thus, the quad wake may be significant in $y$. Note that the effect is smaller by the factor 0.49 if we keep the beam a distance 0.7~mm from one wall, but move the other wall far away (see Eq.~\ref{wq0_wall_eq}).
The dipole kick transforms $y'$ from position $i$ to $f$ as: $y_f'=y_i'+f_d(s)$ with
\begin{equation}
f_d(s)=\left(\frac{Z_0c}{8\pi}\right)\left(\frac{eQL}{E\ell}\right)\bar w_{d0}'s^2\ .\label{fdb_eq}
\end{equation}
(Here $\bar w_{d0}'=\ w_{d0}'$ (Eq.~\ref{wq0_eq}) with $\Delta y_0$ averaged over the transverse bunch distribution of $\Delta y_0$; in the case of a symmetric distribution like a Gaussian, $\bar w_{d0}'=\ w_{d0}'$ with $\Delta y_0=0$.)

Near the axis the inverse focal length (for a short, uniform bunch) is given by
\begin{equation}
f_q^{-1}(s)=k_q^2(s)L=\frac{\pi^3}{256a^4}Z_0c\left(\frac{eQL}{E\ell}\right)s^2\ .\label{fqb_eq}
\end{equation}
The dipole kick transforms $y'$ from position $i$ to $f$ as: $y_f'=y_i'+y_cf^{-1}_d(s)$ with (see Eqs.~\ref{wxy0pexp_eq}, \ref{wq0_eq})
\begin{equation}
f^{-1}_d(s)=\frac{\pi^3}{128a^4}Z_0c\left(\frac{eQL}{E\ell}\right)s^2\ .\label{fdb_eq}
\end{equation}

The projected emittance is defined by the second moments of the transverse distribution, in thin lens approximation ({\it i.e.} it is assumed that the wake kicks only affect $x'$ and $y'$ distributions). The $y$ emittance is given by  
\begin{equation}
\epsilon_y^2=\sigma_{yy}\sigma_{y'y'}-\sigma_{yy'}^2\ ,
\end{equation}
where $\sigma_{yy}=\sigma_{y}^2$ is left unperturbed, and
\begin{equation}
\sigma_{y'y'}=\langle (y')^2\rangle-\langle y'\rangle^2\ ,\quad
\sigma_{yy'}=\langle yy'\rangle-\langle y\rangle\langle y'\rangle\ 
\end{equation}
(the equivalent equations hold also in $x$).
The averaging is performed over a 2D Gaussian distribution in transverse phase space and the uniform distribution in the longitudinal dimension. 

For example, with the beam offset by $y_c$, the distribution in $(y,y')$ is given by
\begin{equation}
\rho(y,y')=\frac{1}{2\pi\epsilon_y}\exp\Bigg[-\frac{1}{2\epsilon_y}\Bigg(\frac{1+\alpha_y^2}{\beta_y}(y-y_c)^2+ 2\alpha_y y' (y-y_c)+\beta_y\left(y'\right)^2\Bigg)\Bigg]\ ,
\end{equation}
with $\alpha_y$, $\beta_y$, the Twiss parameters at the dechirper. When the beam passes the dechirper near the axis, $y'$ transforms from position $i$ to $f$ as ($y$ is unchanged): 
\begin{equation}
y_f'=y_i'+(y-y_c)f^{-1}_q(s)+y_cf^{-1}_d(s)\ . \label{trans_eq} 
\end{equation}
The distributions transform as
\begin{equation}
\rho_i(y_i,y_i')dy_i\,dy_i'=\rho_f(y_f,y_f')dy_f\,dy_f'=\rho_i(y_i,y_i')J(y_i,y_i';y_f,y_f')dy_f\,dy_f'\ ,
\end{equation}
with $J(y_i,y_i';y_f,y_f')$ the Jacobian of the transformation. For the transformation of Eq.~\ref{trans_eq}, $J=1$.
Thus, the final value of $\langle (y')^2\rangle$ is given by
\begin{equation}
\langle (y')^2\rangle=\frac{1}{\ell}\int_0^\ell ds\int_{-\infty}^\infty dy\int_{-\infty}^\infty dy' \,(y')^2\rho\left[y,y'-(y-y_c)f^{-1}_q(s)-y_cf^{-1}_d(s)\right] \ ,
\end{equation}
and similar for the other first and second moments of the final distribution.
Performing all the integrals needed for the emittance, we find, in general, 
\begin{equation}
\left(\frac{\epsilon_y}{\epsilon_{y0}}\right)=\left[1+\beta_y^2\left(\left[f_q^{-1}\right]_{rms}^2+\frac{y_c^2}{\sigma_y^2}\left[f_d^{-1}\right]_{rms}^2\right)\right]^{1/2}\ ,\label{eps_eq}
\end{equation}
with the notation
\begin{equation}
\left[f_q^{-1}\right]_{rms}^2=\int ds \lambda(s)\left[f_q^{-1}(s)\right]^2-\left(\int ds \lambda(s)\left[f_q^{-1}(s)\right]\right)^2\ ,
\end{equation}
where $\lambda(s)$ is the longitudinal bunch distribution (here a uniform longitudinal distribution is not assumed). In the brackets of Eq.~\ref{eps_eq}, after the 1, we see, respectively, the quad and dipole wake contributions to the emittance growth. Note that $\alpha_y$ does not appear in the solution. The same equation holds for $(\epsilon_x/\epsilon_{x0})$ if we replace $\beta_y$ with $\beta_x$ and remove the dipole contribution. 
For the uniform longitudinal distribution and with the beam near the axis, we obtain finally (using Eqs.~\ref{fqb_eq}, \ref{fdb_eq}, \ref{eps_eq})
\begin{equation}
\left(\frac{\epsilon_y}{\epsilon_{y0}}\right)=\left[1+\left(\frac{\pi^3}{384\sqrt{5}}\frac{Z_0c}{a^4}\frac{eQ\beta_y L\ell}{E}\right)^2\left(1+4\frac{y_c^2}{\sigma_y^2}\right)\right]^{1/2}\ ,\label{epsf_eq}
\end{equation}

Applying the parameters of Table~I with $y_c=0$, we obtain projected emittance growths due to just the quad wake: $(\epsilon/\epsilon_0)=1.005$ (1.14) in $x$ ($y$). If we misalign the beam vertically by $y_c=25$~$\mu$m, then $(\epsilon_y/\epsilon_{y0})=1.53$. These results are somewhat larger than given in Ref.~\cite{Zhang15}; we attribute the discrepancy to the fact that the model of the earlier report underestimates the wakes near the origin while our model overestimates them. Also, note that, for emittance growth, $(\epsilon_y-\epsilon_{y0})/\epsilon_{y0}$, small compared to 1, the dipole contribution scales as $\beta_y/\epsilon_{y0}$.

\subsection*{Thick Lens Calculation}

To this point, our emittance calculations have been thin lens calculations. For the dechirper, the effective focusing strength due to the quad wake is given by (see Eq.~\ref{fm1_eq}) 
\begin{equation}
k_q(s)=\left(\frac{\pi^3}{256}{Z_0c}\frac{eQ}{E\ell}\right)^{1/2}\frac{s}{a^2}\ .
\end{equation}
Using the parameters of Table~I, we find that, at the tail of the bunch,
$k_q(\ell)L=0.39$, and one would think that the thin lens approximation may suffice. However, in order to cancel the quad wake effect, a two-meter-long vertical dechirper is followed by a two-meter-long horizontal one. In the thin lens approximation, a vertical quad next to a horizontal quad will completely cancel the focusing and defocusing effects. We perform here thick lens calculations with the proper spacing of the elements to investigate how well the cancellation really works. 

We consider the case with the beam centered on the axis. The quad wake, for any slice position $s$, transforms just like a magnetic quadrupole (see {\it e.g.} Ref.~\cite{Handbook}). We can transform the initial beam ellipse properties in one plane, $T_0=(\beta_0,\alpha_0,\gamma_0)^{T}$, with $\gamma=(1+\alpha^2)/\beta$ and superscript $T$ means the transpose of a matrix, to the final state, $T_f=R(s)T_0R(s)^T$. Then the emittance growth
\begin{equation}
 \left(\epsilon_f/\epsilon_0\right)=\left(\langle\gamma_f\rangle\langle\beta_f\rangle-\langle\alpha_f\rangle^2\right)^{1/2} \ ,\label{epsz_eq} 
\end{equation}
where $\langle\rangle$ means to numerically average (integrate) over $s$. Then, the same is repeated in the other plane.

The matrices for a thick lens focusing and defocusing quad are, respectively, given by
\begin{equation}
R_f=\left(
\begin{array}{cc}
\cos k_qL &\frac{\sin k_qL}{k_q}\\
-k_q\sin k_qL &\cos k_qL\\
\end{array}
\right)\ ,\quad\quad
R_d=\left(
\begin{array}{cc}
\cosh k_qL &\frac{\sinh k_qL}{k_q}\\
k_q\sinh k_qL &\cosh k_qL\\
\end{array}
\right)\ .
\end{equation}
We begin with a single two-meter, horizontal dechirper, where focusing occurs in $x$ and defocusing in $y$. For each of the three averages needed for the emittance (see Eq.~\ref{epsz_eq}), we first analytically performed the matrix multiplications and, at the end, numerically integrated over $s$. We find that, for the thick lens calculation, $(\epsilon/\epsilon_0)= 1.005$ (1.19) in $x$ ($y$), which represents in $x$ nearly the same answer as the thin lens result, and in $y$ a 35\% increase over the thin lens emittance growth.

To see how well the cancellation using the two dechirper sections works we performed a thick lens calculation for the LCLS configuration, consisting of: (1)~a vertical 2-meter dechirper, (2)~a 0.5~m drift, (3)~a thin lens, horizontally focusing quad (with a 7.7~m focal length), (4)~another 0.5~m drift, followed by (5)~a horizontal 2-meter dechirper. For this calculation we first multiplied the $R(s)$ matrices of the five components together, transformed to obtain the final Twiss parameters $T_f(s)$, and then numerically averaged the Twiss parameters over $s$ as described above. The final result is that $(\epsilon/\epsilon_0)=1.008$ (1.12) in $x$ ($y$). We see that by adding a second, rotated dechirper we slightly increased the projected $x$ emittance but, at the same time, significantly reduced the $y$ emittance.

\section*{Accuracy}

For the steady-state component of the wakes to well approximate the total wake, the catch-up distance, $z_{cu}=a^2/2\sigma_z$, must be small compared to the structure length. For the parameters of Table~I, taking $\sigma_z=\ell/(2\sqrt{3})=8.7$~$\mu$m, $z_{cu}=3$~cm, and this condition is clearly satisfied. This means that the longitudinal wake at the origin that we have been using is correct. However, rather than being relatively constant, the wake may already drop somewhat over the short distance of interest.
In the transverse case, the slope at the origin, as given by our model, is correct, but it may also drop.

The original work on round~\cite{BaneNovo99}, \cite{BaneStupakov00}, and flat~\cite{BaneStupakov03} corrugated structure wakes used perturbation methods to obtain the wake functions, which were found to be well approximated by a single cosine (single damped cosine) function in the round (flat) case. The two assumptions for the validity of the results are that: (1)~the depth to period ratio of the corrugations is not small, $h/p\gtrsim1$ (see Fig.~\ref{fig:1}), and (2)~$h/a\ll1$. In practice, it is difficult to manufacture corrugation features that are smaller than $\sim0.5$~mm, and small half gaps $a\sim1$~mm are desirable because they result in a strong dechirping effect; the result is that the validity conditions for the perturbation methods tend not to be satisfied. For example, for the RadiaBeam/LCLS dechirper with the nominal half gap, $a=0.7$~mm, the ratio $h/a=0.7$, which is not small.

In Ref.~\cite{Novo15} A. Novokhatski performed numerical, time-domain studies of the wakes in the RadiaBeam/LCLS dechirper and found that higher order modes can add significantly to the wake near the origin. 
In Fig.~2 of that report, which gives the (on-axis) bunch wake of a $\sigma_z=25$~$\mu$m Gaussian bunch in the RadiaBeam/LCLS dechirper with the half-gap $a=0.7$, one sees a $\sim25\%$ initial amplitude drop to a damped cosine (average) variation. To obtain accurate results for the RadiaBeam/LCLS dechirper, one needs to run a numerical code, such as the time domain code used in ~\cite{Novo15} or ECHO(2D)~\cite{Zagorodnov15}.

We have performed test runs with ECHO(2D) for the parameters of Table~I, using a Gaussian beam with rms length $\sigma_z=10$~$\mu$m, and find good agreement in longitudinal and quad bunch wakes with results presented in~\cite{Novo15}. 
The bunch shape in the LCLS is, however, nominally uniform.
For a uniform bunch distribution of length $\ell$, the loss factor can be obtained from 
\begin{equation}
\varkappa=\frac{1}{\ell^2}\int_0^\ell w(s)(\ell-s)\,ds\ ,
\end{equation}
(and the equivalent equations hold for the kick factors).

In Table~II we summarize the numerical [ECHO(2D)] results for the nominal bunch length ($\ell=30$~$\mu$m), as well as for a bunch of twice this length. We give both the ratios of loss factors $(\varkappa)_a/\varkappa$ and the bunch wakes at the tail of the bunch $[W_\lambda(\ell)]_a/W_\lambda(\ell)$; here, the numerator (denominator) gives the analytical (numerical) result. In summary, for the nominal parameters of Table~I (with the beam on the axis), the analytical model appears to over-estimate the wakes by $\sim20$--30\%; for a bunch twice as long, the error increases by $\sim10\%$.  
 For shorter bunches, the agreement between the model and the numerical results will become better. 

\begin{table}[hbt]
   \centering
   \caption{For bunches with a uniform longitudinal distribution of full length $\ell$, passing through the RadiaBeam/LCLS dechirper on axis: ratio of the loss and quad kick factors for the analytical model and the ECHO(2D) simulations.
The dipole wake results are approximately the same as the quad ones. The half gap $a=0.7$~mm.}
   \begin{tabular}{||c||c|c||c|c||}\hline 
        $\ell$ [$\mu$m] & $(\varkappa)_a/\varkappa$  &$[W_\lambda(\ell)]_a/W_\lambda(\ell)$ & $(\varkappa_{yq})_{a}/\varkappa_{yq}$& $[W_{\lambda yq}(\ell)]_a/W_{\lambda yq}(\ell)$ \\ \hline\hline 
30 & 1.21 &1.27 &1.28 &1.34  \\ \hline\hline 
60 &1.31 &1.40 &1.42 & 1.51 \\     \hline \hline 
   \end{tabular}
   \label{table2_tab}
\end{table}

\section*{Conclusions}

We began with general expressions for the wakefields in a corrugated structure dechiper with flat geometry, derived in~\cite{surface_imped_flat_geometry}. We took the limits of short bunch length and obtained simplified, approximate expressions for the longitudinal and transverse wakefields, functions that are reasonably accurate for the type of bunch lengths used in {\it e.g.} the LCLS. We then used these functions to obtain, for a short bunch passing through a
dechirper: the energy loss as function of gap, the transverse kick as function of beam offset, the slice energy spread, and the emittance growth of the beam. We performed a thick lens calculation of emittance growth for the two-dechirper system applied to representative LCLS bunch and machine parameters and found that the cancellation does indeed work; the final projected emittance growth is modest: $(\epsilon/\epsilon_0)=1.008$ (1.12) in $x$ ($y$). 

We briefly investigated the accuracy of our model, and find that it overestimates the wakes by $\sim 20$--30\% for the type of parameters considered in this study (in particular, full bunch length $\ell=30$~$\mu$m and dechirper half-gap $a=0.7$~mm and the bunch on axis). For shorter bunches the model becomes more accurate. 

In the Appendix we perform a more accurate derivation of the general wake solution for the dechirper (the starting point of the main work of this report) than found in Ref.~\cite{surface_imped_flat_geometry}. We find that the results of \cite{surface_imped_flat_geometry} are valid provided that the bunch length over half gap, $\sigma_z/a$, is small, which is indeed the parameter regime of interest.

Although in our example calculation the emittance increase was modest, and the cancellation of emittance growth using the two dechirper sections was effective, we would like to point out that large emittance growth is not far away in parameter space, especially if the gap is decreased (note the $a^{-4}$ dependence in Eq.~\ref{epsf_eq}). For example, if in the example calculation above we reduce the energy from $E=6.6$~GeV to $4$~GeV, and the half gap $a=0.7$~mm to 0.5~mm, we find that $k_qL=1.0$. For this example we obtain a large emttance growth for the full, two dechirper, thick lens calculation: $(\epsilon/\epsilon_0)=1.5$ (2.5) in $x$ ($y$).

Finally, we would like to emphasize that although we started with a surface impedance, the surface impedance itself never is part of the solution. This means that the results given here are valid not only for the corrugated pipe dechirper, but rather for {\it any} structure with flat geometry where the impedance can be described by a surface impedance. Such problems include the resistive wall, the shallow corrugated structure (often used as a model for surface roughness), and a metallic pipe lined with a thin dielectric layer. The only condition for their applicability is that the bunch lengths of interest are small compared to the characteristic distance of the problem.

\section*{Acknowledgements}

We thank: the team commissioning the RadiaBeam/LCLS dechirper, led by R. Iverson, for showing us how the dechirper affects the beam in practice; J. Zemella, who has worked with us and performed numerical calculations of the dechirper wakes; I. Zagorodnov, for making his computer code ECHO(2D) available and helping us with its use. Work supported by the U.S.
Department of Energy, Office of Science, Office of Basic
Energy Sciences, under Contract No. DE-AC02-76SF00515.

\appendix
\section{Derivation of the longitudinal impedance with the surface impedance boundary conditions at parallel plates}

 We begin from Maxwell's equations in which we assume that all quantities depend on time and $z$ as $e^{-i\omega t+ikz}$ with $k=\omega/c$ and make the Fourier transform over $x$,
    \begin{align}\label{eq:9}
    \hat f(q)
    =
    \int_{-\infty}^\infty
    dx
    f(x)
    e^{iqx}
    ,\qquad
    f(x)
    =
    \frac{1}{2\pi}
    \int_{-\infty}^\infty
    dq
    \hat f(q)
    e^{-iqx}
    ,
    \end{align}
where $f$ denotes a component of the electromagnetic field. To simplify the notation, in what follows, we drop the hats in the Fourier transformed components of the electromagnetic field. 

The trajectory of the driving particle has a zero horizontal offset, $x=0$, but it is offset in the vertical direction, $y=y_0$. The current density corresponding to the the driving particle in $\omega$ representation is $j_z = I_\omega\delta(x)\delta(y-y_0)e^{-i\omega t+ikz}$ where $I_\omega$ is the amplitude of the Fourier component of the current. The electromagnetic field generated by the driving particle satisfy the Maxwell equations in free space: 
    \begin{align}\label{eq:10}
    \frac{\p E_z}{\p y}
    &=
    -iq
    H_z
    ,
    \nonumber\\
    \frac{\p H_z}{\p y}
    &=
    iq
    E_z
    ,
    \nonumber\\
    E_x
    &=
    H_y
    -
    \frac{q}{k}
    E_z
    ,
    \nonumber\\
    H_x
    &=
    -
    E_y
    -
    \frac{q}{k}
    H_z
    ,
    \nonumber\\
    -iqE_y
    -
    \frac{\p H_y}{\p y}
    &=
    iH_z
    \left(
    k
    +
    \frac{q^2}{k}    
    \right)
    ,
    \nonumber\\
    -iq H_y
    +
    \frac{\p
    E_y}{\p y}
    &=
    -
    i
    E_z
    \left(
    k
    +
    \frac{q^2}{k}
    \right)
    +
    \frac{4\pi}{c}
    I_\omega
    \delta(y-y_0)
    .
    \end{align}
The last term on the right-hand side accounts for the current associated with the particle. Given that the solution of these equation is proportional to the current $I_\omega$, to simplify notation, below we set $I_\omega = 1/2\pi$; this makes the last term on the right-hand side of the last equation in~\eqref{eq:10} equal to $2\delta(y-y_0)$.

The Maxwell equations are supplemented by the boundary conditions at the upper and lower walls,
    \begin{align}\label{eq:40}
    E_z|_{y=a}
    &=
    \zeta H_x|_{y=a}
    ,\qquad
    E_x|_{y=a}
    =
    -
    \zeta H_z|_{y=a}
    ,\nonumber\\
    E_z|_{y=-a}
    &=
    -\zeta H_x|_{y=-a}
    ,\qquad
    E_x|_{y=-a}
    =
    \zeta H_z|_{y=-a}
    ,
    \end{align}
where $\zeta$ is the surface impedance. We seek $H_y$ and $E_y$ in the following form
    \begin{align}\label{eq:41}
    H_y(y)
    =
    A_1 \cosh (q (y-y_0))
    +
    A_2 \sinh (q (y-y_0))
    +
    i\,\mathrm{sign}(y-y_0) \sinh (q(y-y_0))
    ,
    \nonumber\\
    E_y(y)
    =
    B_1 \sinh (q (y-y_0))
    +
    B_2 \cosh (q (y-y_0))
    +
    \mathrm{sign}(y-y_0) \cosh (q(y-y_0))
    ,
    \end{align}
where the terms with $\mathrm{sign}(y-y_0)$ are due to the presence of the delta function source in~\eqref{eq:10} and the amplitudes $A_1$, $B_1$, $A_2$ and $B_2$ are arbitrary numbers. Substituting these equations into the last equation of~\eqref{eq:10} we find the longitudinal electric field
    \begin{align}\label{eq:41}
    E_z
    =
    \frac{iqk}{k^2+q^2}
    (iA_1+B_1)
    \cosh (q (y-y_0))
    +
    \frac{iqk}{k^2+q^2}
    (iA_2+B_2)
    \sinh (q (y-y_0))
    .
    \end{align}
Similarly, $E_x$, $E_y$ and $H_z$ can be expressed through $A_1$, $B_1$, $A_2$ and $B_2$ and the hyperbolic functions from the third, fourth and the fifth equations of~\eqref{eq:10}. After that the amplitudes $A_1$, $B_1$, $A_2$ and $B_2$ can be found from the four boundary conditions~\eqref{eq:40}. Substituting them in~\eqref{eq:41} we obtain the following result,
    \begin{align}\label{eq:42}
    E_z
    =
    \zeta  k q
    \frac{N}{D}
    \end{align}
where    
    \begin{align}\label{eq:43}
    N
    &=
    -i \zeta  k^2
    \sinh \left(q \left(2
    a-y-y_0\right)\right)
    -
    i \zeta  k^2
    \sinh \left(q \left(2
    a+y+y_0\right)\right)
    \nonumber\\
    &+
    \left(\zeta
    ^2+1\right) k q \cosh \left(q\left(2 a-y-y_0\right)\right)
    -
    2 \left(1-\zeta^2\right) k q 
    \cosh \left(q\left(y-y_0\right)\right)
    \nonumber\\&
    +
    k q \cosh\left(q \left(2a+y+y_0\right)\right)
    +
    \zeta^2 k q \cosh \left(q \left(2a+y+y_0\right)\right)
    \nonumber\\&
    +
    i \zeta  q^2\sinh \left(q \left(2a-y-y_0\right)\right)
    +
    i \zeta  q^2
    \sinh \left(q \left(2a+y+y_0\right)\right)
    \end{align}
and
    \begin{align}\label{eq:44}
    D
    &=
    \left(\sinh (2 a q) \left(\zeta ^2
    k^2-q^2\right)+2 i \zeta  k q \cosh
    (2 a q)\right) 
    \nonumber\\&
    \times
    \left(\sinh (2 a q)
    \left(k^2-\zeta ^2 q^2\right)+2 i
    \zeta  k q \cosh (2 a q)\right)
    .
    \end{align}
The longitudinal impedance is related to $E_z$ by the following formula
    \begin{align}\label{eq:45}
    Z_l
    =
    -\frac{1}{I_\omega}
    \frac{1}{2\pi}
    \int_{-\infty}^\infty
    E_z(q)
    e^{-iqx}
    =
    -
    \int_{-\infty}^\infty
    E_z(q)
    e^{-iqx}
    ,
    \end{align}
where in the last equality we recalled our assumption $I_\omega=1/2\pi$. Substituting~\eqref{eq:42} into Eq.~\eqref{eq:44} gives a general expression for the longitudinal impedance without any assumption. The expression~\eqref{tildeZl2_eq} that we use in the main body of this paper is obtained from ~\eqref{eq:42} and~\eqref{eq:44} if one takes into account that $|\zeta|\ll 1$ and also assumes $q\ll k$. The latter is justified for short bunches with $\sigma_z\ll a$. With these assumptions, we can neglect the last three terms in Eq.~\eqref{eq:43}, replace $1\pm\zeta^2\to 1$, and neglect the term $\zeta^2q^2$ in Eq.~\eqref{eq:44}. It is easy to check that with these modifications impedance~\eqref{eq:44} is the same as Eq.~\eqref{tildeZl2_eq}.

                            
\end{document}